\documentclass[journal]{IEEEtran}

\ifCLASSINFOpdf
\else
   \usepackage[dvips]{graphicx}
\fi
\usepackage{url}

\usepackage{amsmath}
\usepackage{amssymb}
\usepackage{graphicx}
\usepackage{cite}
\usepackage{algorithm}
\usepackage{algpseudocode}
\usepackage{mathtools}

\DeclareMathOperator{\E}{\mathbb{E}}
\DeclareMathOperator{\x}{\mathbf{x}}
\DeclareMathOperator{\bmu}{\boldsymbol{\mu}}

\begin{document}

\bibliographystyle{IEEEtran}

\title{Rare Events via Cross-Entropy Population Monte Carlo}

\author{Caleb Miller, Jem N. Corcoran, and Michael D. Schneider
\thanks{``This work was performed under the auspices of the U.S. Department of Energy by Lawrence Livermore National Laboratory under Contract DE-AC52-07NA27344. Funding for this work was provided under LLNL Laboratory Directed Research and Development 19-SI-004.'' }
\thanks{C. Miller and J. Corcoran are with Applied Mathematics Department, University of Colorado, Boulder (e-mail: caleb.miller@colorado.edu, corcoran@colorado.edu)}
\thanks{Michael Schneider is a group leader in physics division of LLNL (email: schneider42@llnl.gov)}}

%\markboth{IEEE SIGNAL PROCESSING LETTERS, VOL. XX, NO. X, XXXX}
%{Shell \MakeLowercase{\textit{et al.}}: Bare Demo of IEEEtran.cls for IEEE Journals}
\maketitle

\begin{abstract}
We present a Cross-Entropy based population Monte Carlo algorithm. This methods stands apart from previous work in that we are not optimizing a mixture distribution. Instead, we leverage deterministic mixture weights and optimize the distributions individually through a reinterpretation of the typical derivation of the cross-entropy method. Demonstrations on numerical examples show that the algorithm can outperform existing resampling population Monte Carlo methods, especially for higher-dimensional problems.
\end{abstract}


\IEEEpeerreviewmaketitle

\section{Introduction}

\IEEEPARstart{R}{are} events are events that happen with very low frequency. While the definition of low frequency is domain specific, the terminology is typically reserved for events deemed disruptive and even catastrophic, in areas such as such as in structural reliability \cite{kurtz2013cross}, conjunction assessment \cite{losacco2019advanced}, climate modeling \cite{webber2019}, and epidemiology \cite{machado2020}. Estimating rare event probabilities using Monte Carlo techniques is computationally expensive, often to the point of intractability, and special techniques are required \cite{rubino2009collection}. Such techniques include  subset splitting, line sampling, and  importance sampling. In the case of importance sampling, a proposal distribution must be chosen by the user, and a poor choice will have the undesirable effect of producing high variance estimates. Adaptive importance sampling (AIS) algorithms allows one to update the proposal distribution based on intermediate results. A well known AIS algorithm is the cross-entropy algorithm of Rubenstein and Melamed \cite{rubinstein1998,rubinstein2013cross} which aims to minimize the  Kullback–Leibler divergence between the proposal distribution and the optimal sampling density from a given parametric family using incremental parameter changes. 

The population Monte Carlo (PMC) algorithm \cite{cappe2004population} is an AIS algorithm that can be used for estimating rare event probabilities or, more generally, expectations with respect to a given target distribution. At each iteration, a Markov transition kernel is used to propagate a set of particles (samples) forward in time. Importance sampling weights are attached to each propagated particle and a new set of particles is sampled according to those weights. At any stage, the weighted particles can be used to give an unbiased estimator of the probability of interest. In fact, all particles and weights from all iterations can be used. In this paper we give brief background on rare events, importance sampling, and the cross-entropy method. We then show how a reinterpretation of the derivation of the method leads naturally to a population Monte Carlo scheme and demonstrate its efficacy on several examples. 

\section{Rare Events and Importance Sampling}

\subsection{Rare Event Problem}
The rare event problem involves estimating probabilities of a random variable $\x$ exceeding a level $\gamma$ of a performance function $S(\x)$. Assuming $\x$ is distributed according to a probability distribution function $\pi(\x)$ we are seeking to estimate \[\ell = \mathbb{P}(S(\x)\geq \gamma) = \E_\pi\left[ I_{\{S(\x)\geq \gamma\}}\right],\] where $I$ is an indicator function. Due to the rareness of the event standard Monte Carlo techniques require exorbitant amounts of samples to provide accurate estimates. To improve on this importance sampling techniques are often applied to find important regions, that is, regions where samples are more likely to exceed the required performance level. 

\subsection{Importance Sampling}
Importance sampling (IS) is a ubiquitous Monte Carlo technique that draws samples from a proposal distribution and weights them accordingly to match expectations from a target distribution \cite{robert2013monte}. In terms of the rare event problem and given a proposal distribution with pdf $q(\x)$ the IS estimator is

\[\E_\pi\left[ I_{\{S(\x)\geq \gamma\}}\right] \approx \hat{\ell} = \frac{1}{K}\sum_{k=1}^K I_{\{S(\x_k)\geq \gamma\}} w(\x_k),\] where $\x_k$ are drawn from the importance distribution $q$ and $w(\mathbf{x})_k = \pi(\x_k)/q(\x_k)$ are the importance weights. The importance weights can be interpreted as the Radon-Nikodym derivative between the measures that correspond to the pdfs $\pi$ and $q$. Central to the effectiveness of IS is the choice of the proposal distribution. Naturally, many methods have been employed to iteratively adapt the proposal distribution. Furthermore, rather than adapting a single distribution, adapting a population of distributions, $q_1,\cdots,q_N$, has spurred many AIS methods which are well-reviewed by Bugallo et al. \cite{bugallo2017adaptive}. An important advance in the weighting scheme of multiple importance sampling (MIS) is the concept of the deterministic mixture weight (DM-weight) \cite{elvira2019generalized}. In this weighting scheme the samples are weighted as if drawn from the mixture of distributions, that is the importance weight for a sample is given by
\begin{equation} \label{DM}
    w(\x_k) = w_k =  \frac{\pi(\x_k)}{\frac{1}{N}\sum_{n=1}^N q_n(\x_k)},
\end{equation}
where we have introduced the notation $w_k$ to mean the weight of the $k$th sample.
The advantage of this weighting scheme is it reduces the variance of the IS estimators. 

Several variations of the adapting the family of proposals have been proposed. Adaptation through resampling schemes: local resample PMC (LR-PMC) adapts a proposal in the family through multinomial resampling based on samples produced from that proposal for all proposals independently, in contrast global resample PMC (GR-PMC) adapts all proposals at once by performing multinomial sampling on all samples produced by all the proposals \cite{elvira2017population}. Gradient adaptive population importance sampler (GAPIS) utilizes the target distributions gradient and Hessian matrix to adapt proposals \cite{elvira2015gradient}. Utilizing popular Markov chain Monte Carlo (MCMC) techniques has also been studied recently to adapt proposals e.g. Langevin based PMC (SL-PMC) \cite{elvira2019langevin} and Hamiltonian AIS (HAIS) \cite{mousavi2021hamiltonian}. Further MCMC-driven IS techniques are discussed by Llorent et al. \cite{llorente2021mcmc}. An important realization discussed by Llorent et al. is that it is unknown what distribution the MCMC methods should target. One way of producing the best possible distribution to perform importance sampling from is the cross-entropy method \cite{rubinstein2013cross}. Here, the measure of best is the Kullback-Leibler divergence between the optimal importance sampling distribution and the proposal distributions. Mixture distributions have previously been adapted via the cross-entropy process \cite{kurtz2013cross,wang2016cross}, and through an expectation-maximization like procedure in mixture PMC \cite{cappe2008adaptive} and D-kernel PMC \cite{douc2007convergence}. While the method in this paper utilizes the mixture weights we focus on adapting individual proposals as opposed to the mixture of distributions, this eliminates update equations for the weights of the mixture. As discussed below, the method in this paper involves gradients with respect to the parameters of the proposal distributions, which stands apart from GAPIS, SL-PMC, and HAIS which require gradients of the logarithm of the target distribution. This is especially helpful in the case of rare events which are often posed in the form of satisfying a performance or indicator function, which would necessarily create discontinuities in the gradient of the target distribution.

\section{Cross-Entropy Method}
The cross-entropy method is often utilized for rare-events and combinatorial optimization tasks. In particular we utilize the multi-level algorithm as described by DeBoer et al. \cite{de2005tutorial}. An importance sampling proposal distribution, indexed by  parameters $\bmu$, is selected, and the optimal parameter is sought through incremental changes.  The algorithm proceeds in two stages, involving updating temporary performance levels $\hat{\gamma_t}$ to build up to the desired performance $\gamma$ and then updating the parameters of the proposal from $\bmu^{(t)}$ to $\bmu^{(t+1)}$. From step $t$ to step $t+1$ of the algorithm, samples $\{\x_k\}_{k=1}^K$ are obtained from the proposal distribution $q(\cdot ;\bmu^{(t)})$, the performance function is evaluated on the samples, these performances are then sorted and the $(1-\rho)$ sample quantile of the performances is used to determine the temporary performance level. Samples that provide performance beyond the temporary performance level are then used to update the parameters to minimize the Kullback Leibler divergence between the optimal sampling density and proposal distribution, this results in optimization problem

\begin{equation}
\label{ce1}
\max_{\bmu} \frac{1}{K} \sum_{k=1}^K I_{\{S(\x_k)\geq \hat{\gamma}_t\}} \frac{\pi(\x_k)}{q(\x_k;\bmu^{(t)})}\ln q(\x_k;\bmu).
\end{equation}

%When updating a single Gaussian distribution mean $\bmu$ and covariance $\bsig$ the update equations are given by
%\begin{align*}
%\label{e:update}
%    \bmu &= \frac{\sum_{k=1}^K I_{\{S(\x_k)\geq \hat{\gamma}_t\}} w_k %\x_k}{\sum_{k=1}^K I_{\{S(\x_k)\geq \hat{\gamma}_t\}} w_k} \\
%    \bsig &= \frac{\sum_{k=1}^K I_{\{S(\x_k)\geq \hat{\gamma}_t\}} w_k %(\x_k-\bmu)(\x_k-\bmu)^T}{\sum_{k=1}^K I_{\{S(\x_k)\geq \hat{\gamma}_t\}} w_k}.
%\end{align*}

\begin{algorithm}[H]
\caption{Cross-Entropy Algorithm}\label{CE}
\begin{algorithmic}[1]
\State \textbf{Input}: Quantile parameter $\rho$, initial importance sampling parameter $\bmu^{(0)}$, number of values to be sampled $K$
\State Set $t=0$ and set $\hat{\gamma}_{0}$ to be any value below $\gamma$.
\While{$\hat{\gamma}_{t}<\gamma$}
\State Set $t = t+1$.
\State \textbf{sampling:} Draw $K$ samples from the importance sampling density  \[\{\x_{k}^{(t)}\}_{k=1}^K \sim q(\cdot ;\bmu^{(t-1)}).\]
    \State \textbf{performances:} Evaluate the performance function $S(\x)$ at each sampled value and order the results to produce   $S_{(1)} \leq S_{(2)} \leq \cdots \leq S_{(K)}$.
    \State \textbf{sample quantile}: Let  
    $\hat{\gamma}_t = S_{\lceil (1-\rho)K \rceil}$ be the $(1-\rho)$ sample quantile of the performances.
    \State \textbf{if} $\hat{\gamma}_t<\gamma$ \textbf{then}
     \State \indent Solve the CE Update Equation \eqref{ce1} to obtain $\bmu^{(t)}$.

     \State \textbf{else}
    \State \indent Set $\hat{\gamma}_t =\gamma$.
    \State \textbf{end if}

\EndWhile

\State \textbf{Output}: $\hat{\ell} = \frac{1}{K} \sum_{k=1}^{K} I_{\{S(x_{k}^{(t)}) \geq \gamma\}} \frac{\pi(x_{k}^{(t)})}{q(x_{k}^{(t)},\bmu^{(t)})}$.
\end{algorithmic}
\end{algorithm}

%\begin{algorithm}[H]
%\caption{Cross Entropy Algorithm}\label{CE}
%\begin{algorithmic}[1]
%\State \textbf{Input}: Samples $\{\x_{k}\}_{k=1}^K$, Weights %$\{w(\x_{k})\}_{k=1}^K$ , quantile parameter $\rho$
%    \State \textbf{Performances} Evaluate and order sample performances   $S_{(1)},\cdots, S_{(K)}$
%    \State \textbf{Obtain Temporary Level}: 
%    $\hat{\gamma}_t = S_{\lceil (1-\rho)K \rceil}$
%    \State \textbf{Solve CE Update Equation}: to obtain $\bmu^{(t)}$ 
%\State \textbf{Output}: New parameter $\bmu$
%\end{algorithmic}
%\end{algorithm}

\section{Cross-Entropy Population Monte Carlo}
To incorporate the cross-entropy method as a way to update the parameters for the population of proposal distributions, we modify the typical derivation of the cross-entropy method, starting with the  minimization of Kullback-Leibler divergence between the optimal distribution $p(x) = \frac{1}{\ell}I_{(S(\x)\geq\gamma)}\pi(\x)$ to create a new stochastic program to optimize the family of proposals , briefly let $Q_{\bmu}$ denote the family of proposals as a mixture distribution,

\begin{align*}
    &\min_{\bmu} KL\left(p||Q_{\bmu}\right)  = \max_{\bmu} \int p(\x)\ln\left(Q(\x;\bmu)\right)\,d\x \\
    & \approx \max_{\bmu} \frac{1}{NK}\sum_{n=1}^N \sum_{k=1}^K \frac{I_{(S(\x_{n,k})\geq\gamma)}\pi(\x_{n,k})}{\frac{1}{N}\sum_{n=1}^N q_n(\x_{n,k};\nu_n)} \ln\left(Q(\x_{n,k};\bmu)\right) \\
    &= \max_{\bmu} \frac{1}{NK}\sum_{n=1}^N \sum_{k=1}^K \frac{I_{(S(\x_{n,k})\geq\gamma)}\pi(\x_{n,k})}{\frac{1}{N}\sum_{n=1}^N q_n(\x_{n,k};\nu_n)} \ln\left(q_n(\x_{n,k};\bmu_n)\right)
\end{align*}

The first approximation is the unbiased multiple importance sampling approximation with the DM-weights taken from the previous trial. The first equality comes from knowing that the $n$th sample is drawn from the $n$th proposal distribution, so the pdf applied to that sample is $q_n$. Thus the stochastic program splits into $N$ individual optimization problems -- one optimization problem for each proposal distribution. That is for $n=1,\dots,N$ do the following optimization

\begin{equation}
\label{sp}
    \max_{\bmu_n} \frac{1}{K} \sum_{k=1}^K \frac{I_{(S(\x_{n,k})\geq\hat{\gamma}_t)}\pi(\x_{n,k})}{\frac{1}{N}\sum_{n=1}^N q(\x_{n,k};\bmu_n^{(t)})} \ln\left(q_n(\x_{n,k};\bmu_n)\right).
\end{equation}

Intuitively, the parameters of every proposal distribution are updated in a cross-entropy multilevel fashion with the DM-weights in place of the the typical IS weights. Heuristically, the denominator in the DM-weights promotes distance between proposals distributions as samples from a particular proposal distribution will be weighted more heavily when further away from the other proposals. In the algorithm below, we adopt the typical set-up for most PMC methods-- that is we set a number of proposals $N$, number of samples $K$, and number of trials $T$. Fixing the number of samples and trials is not required of the CE-PMC method. In some applications of the cross-entropy method rounds of presampling with a smaller number of samples are used to find the importance region followed by a final trial with a larger number of samples. Instead of a set number of trials, many stopping criteria could be employed e.g the one used in Algorithm \ref{CE}--stopping when the sample quantile exceeds the desired performance.

\begin{algorithm}[H]\label{CEPMC}
\caption{Cross-Entropy Population Monte Carlo}\label{ALGCEPMC}
\begin{algorithmic}[1]
\State \textbf{Input}: Quantile parameter $\rho$, number of proposals $N$, number of samples per proposal $K$, number of trials $T$, and parameters of initial distributions $\{\bmu^{(1)}_n \}_{n=1}^N$ .

    \For{$t=1,\cdots, T$}
        \State \textbf{sampling:} For $n=1,\dots,N$ draw $K$ samples from each proposal \[\{\x_{n,k}^{(t)}\}_{k=1}^K \sim q_n(\cdot ;\bmu_n^{(t)}).\]
        
        \State \textbf{weighting:} Weight every sample with the DM-weight 
        \[ w_{n,k}^{(t)}= \frac{\pi(\x_{n,k}^{(t)})}{\frac{1}{N} \sum_{n=1}^N q_n(\x_{n,k}^{(t)};\bmu_n^{(t)})}. \]
        
        \State \textbf{performances:} Obtain $S_{(i)}$ see Algorithm \ref{CE}. 
        \State \textbf{sample quantile:} Obtain $\hat{\gamma}_t$, see Algorithm \ref{CE}.
        \State \textbf{adapt:} Solve CE Update \eqref{sp}  to obtain $\bmu_n^{(t+1)}$ 
        
    \EndFor
    \State \textbf{Output:} All samples and their respective weights \[\{\x_{n,k}^{(t)}, w_{n,k}^{(t)}\}_{n=1,k=1,t=1}^{N,K,T}\] 
    
\end{algorithmic}
\end{algorithm}

In AIS methods typically the final estimator is given by
\[\hat{\ell} = \frac{1}{TNK} \sum_{t=1}^T \sum_{n=1}^N \sum_{k=1}^K I_{S(\x_{n,k}^{(t)}\geq \gamma)} w_{n,k}^{(t)}.\] As the cross-entropy method should converge to the optimal parameters, in this paper we will only consider the estimate given by the samples produced in the final trial as in the output of Alg. 1.
\section{Numerical Examples}
For each experiment we will adapt a family of Gaussian distributions and compare the LR-PMC, GR-PMC, and CE-PMC methods. Although typically, cross-entropy updates are run until a threshold is met, for a fair and consistent evaluation between the three methods, we fix the number of trials $T$, the number of proposals $N$, and the number of samples  per proposal $K$ for each example. Furthermore, we impose two ad hoc conventions. Proposals can often produce sets of samples all with zero weight, thus halting any effort to perform a multinomial resample. For LR-PMC the convention when a proposal produced all $K$ samples with zero weight, was to reweight the samples evenly with weight $1/K$, and proceed with the algorithm. For GR-PMC, the multinomial resampling breaks down if all $NK$ samples have zero weight, in which case we reweight all the samples evenly with weight $1/NK$ and proceed as the algorithm intended. In CE-PMC when updating the covariances of a Gaussian distribution one or several dimensions may flatten resulting in a singular matrix, especially when approaching a linear function \cite[Sec. 6.3]{geyer2019cross}. We implement two operational procedures to mitigate this problem, the first is to check if the updated covariance matrix is singular, if it is we adopt the covariance from the previous trial, the second procedure is to only utilize the mean updating formula in the first half of the trials, and update both the mean and covariance in the second half of the trials, this is a version of scheduling covariance updates \cite[Sec. 3]{cappe2004population}. More advanced techniques may be implemented to fix this problem, for instance, the modified metropolis algorithm of subset simulation \cite{au2001estimation} could be used to expand the covariance in the directions of decay.

\subsection{Structural Reliability Examples}
 First we examine three examples taken from structural reliability literature \cite{kurtz2013cross,geyer2019cross}. The target distributions are all proportional to $I_{\{S_i(\x) \leq \gamma\}}\pi(\x)$, where $\pi(\x) = \pi(x_1,x_2)$ is given by a standard multivariate Gaussian distribution, and $ S_1(\x)= 5 - x_2 - 0.5(x_1-0.1)$, $S_2(\x)= 5 - x_2 - 0.1(x_1)$ and $S_3(x)$ is a minimum of the following expressions $3+(x_1-x_2)^2/10\pm(x_1+x_2)/\sqrt{2}$ and $\pm(x_1-x_2)+7/\sqrt{2}$.
 
 %\begin{align*}
 %    S_1(\x)  &= 5 - x_2 - 0.5(x_1-0.1), \\
 %    S_2(\x)  &= 5 - x_2 - 0.1(x_1), \\
 %    S_3(\x) & = \text{min} \left\{ 
 %    \begin{aligned}
 %        & 3+(x_1-x_2)^2/10-(x_1+x_2)/\sqrt{2} \\
 %        & 3+(x_1-x_2)^2/10+(x_1+x_2)/ \sqrt{2} \\
 %        & x_1-x_2+7/\sqrt{2} \\
 %        & x_2-x_1+7/\sqrt{2}
 %    \end{aligned}
 %    \right\}.
 %\end{align*}
 
 We refer to the problems respectively as S1, S2, and S3 the problems have the respective reference values 3.01e-3, 8.67e-7, and 2.22e-3. We compare the performance of LR-PMC, GR-PMC, and CE-PMC on these three examples, with $N=25$, $K=100$, and $T=20$ and average the results over 1000 runs. As shown in Table \ref{SR}, the CE-PMC outperformed the other methods with respect to relative root mean squared error (RRMSE). In Fig. 1 we see that the CE-PMC algorithm is able to match complex regions of importance to produce reliable estimates.
 
 \begin{table}[!t]
\renewcommand{\arraystretch}{1.3}
\caption{RRMSE on Structural Reliability Examples}
\label{SR}
\centering
\begin{tabular}{c|c|c|c}
\hline
\bfseries Method/Problem & \bfseries $S1$ &  $S2$  &  $S3$ \\
\hline\hline
LR-PMC  & 0.0424          & 0.0602           & 0.0542 \\
GR-PMC  & 0.0602          & 0.0494           & 0.6603 \\
%CE1-PMC & \textbf{0.0096} & 0.0442           & \textbf{0.0147} \\
CE-PMC & \textbf{0.0163}          & \textbf{0.0141}  & \textbf{0.0233} \\
\hline
\end{tabular}
\end{table}

\begin{figure}
\label{SRfig}
\centerline{\includegraphics[scale=0.4]{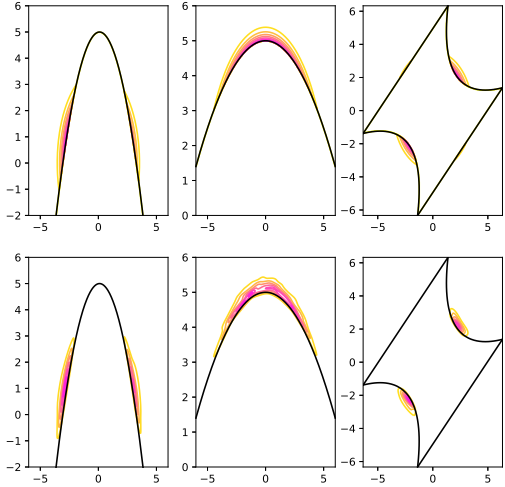}}
%\centerline{\includegraphics[width=\columnwidth]{Figures/Line_Dim_Comp3.eps}}
%\centerline{\includegraphics[scale=0.5]{Figures/Line_Dim_Comp.eps}}
\caption{Top: Contour plots of the desired distribution for S1 (left), S2 (middle), and S3 (right). Bottom: Example of contour plots of the estimated distribution produced by CE-PMC.}
\end{figure}
 
\subsection{Variable Dimension Problem}
 We consider the problem $S_4(x) = \beta - \frac{1}{\sqrt{D}}\sum_{i=1}^D x_i$ where $x_i$ are drawn from standard normal distributions \cite{kurtz2013cross}. This problem is particularly interesting as the probability is $\Phi(-\beta)$ regardless of dimension $D$, where $\Phi$ is the cumulative distribution function of the standard normal. This allows for testing the methods response to changes in dimension. We set $\beta = 5$ so that the true rare event probability is $\Phi(-5)\approx$ 2.86e-7
 and run $100$ independent tests with $N=4, K=5000, T=32$ for each dimensions 2, 5, 10, 20, 30, 40, and 50. The initial means of the distributions were chosen by scaling, between $-1$ and $1$, centered Latin hypercube samples i.e. every entry of $\bmu^{(0)}_n$ is either $\pm 0.25$ or $\pm 0.75$. The initial covariances were isotropic, $\Sigma^{(0)}_n = \sigma^2 \textbf{I}_D$ with $\sigma=1$. The results of this experiment are displayed in the Fig. 2. It is clear that CE-PMC performs better than LR-PMC and GR-PMC as the dimension increases. 
 
 \subsection{Conjunction Analysis}
 A problem of much importance in the space domain community is that of conjunction analysis, which involves finding the probability of an object in space passing nearby another object e.g. satellites and space debris. Consider a rogue object with isotropic Gaussian uncertainty (5 meters in position and 0.1m/s in velocity) at time $t_0$ (denoted the distribution by $\pi$) and Cartesian position denoted by $r$, we are concerned with the probability that this rogue object comes with in $50$ meters at time $t_1 \approx t_0 + 9893.34s  $ of two assets with position denoted by $a_1$ and $a_2$ which are on the same orbital plane $10,000$ meters apart at $t_0$. We simulate the positions and velocities of all objects with Keplerian propagators. Our performance function is $S(x) = \min_i||r-a_i||_2$ and our goal is to estimate the probability that the rogue object comes within $50$ meters of the assets, $\ell = \E_\pi\left[ I_{\{S(\x)\geq 50\,m\}}\right]$. Performing the CE-PMC algorithm with $N=16$ Gaussian distributions, $K=500$ samples per distribution, and $T=20$ trials we estimate $\hat{\ell}_{CE} \approx 0.000113$. To confirm the validity of this result we simulate with 1 million Monte Carlo samples and get the estimate $\hat{\ell}_{MC} = 0.000111$. In Fig. 3 we again see the ability of CE-PMC to find the regions of importance, emphasized by the swath of low performance Monte Carlo samples. 
 
\begin{figure}
\label{Linefig}
\centerline{\includegraphics[scale=0.5]{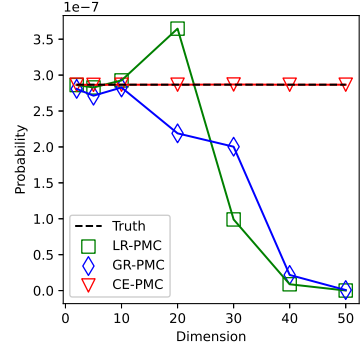}}
%\centerline{\includegraphics[width=\columnwidth]{Figures/Line_Dim_Comp3.eps}}
%\centerline{\includegraphics[scale=0.5]{Figures/Line_Dim_Comp.eps}}
\vspace*{-2mm}
\caption{Mean Probability over 100 Monte Carlo runs against increasing dimension $D$.}
\end{figure}

\begin{figure}
\label{hoco}
\vspace*{-5mm}
\centerline{\includegraphics[scale=0.33]{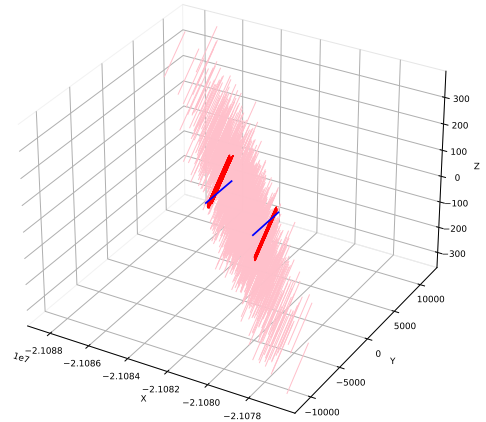}}
%\centerline{\includegraphics[width=\columnwidth]{Figures/Line_Dim_Comp3.eps}}
%\centerline{\includegraphics[scale=0.5]{Figures/Line_Dim_Comp.eps}}
\vspace*{-2mm}
\caption{Traces of orbital position plotted from $t_1-5$ to $t_1+5$: assests (blue), 1000 Monte Carlo samples of the rogue object (light pink), and 50 samples of the rogue object (red) from each of the proposal distributions in the final trail of CE-PMC.}
\end{figure}

\section{Conclusions}
When running experiments the schedule of updating the means in the first half of trials, followed by means and covariances in the following trials, was important to the performance of the algorithm, without this trick the high-dimensional examples proved to be very challenging. Future explorations should consider how smoothing the performance functions could allow for increased performance of CE-PMC and the ability of gradient-based MCMC methods such as HAIS to be more applicable. There are interesting paths of future work in the area of path integral optimal control. Kappen et al.\cite{kappen2016adaptive} have devised formulas for optimal distributions of trajectories and applied an importance sampling cross-entropy method to return optimal controls. Greater exploration of the parameter space of the control could be achieved via CE-PMC. %Another avenue is the data fusion task, treating a product of likelihoods as a rare event performance function, one could run the CE-PMC algorithm to gain insights about parameters of a desired target. 

\newpage

%\section*{Acknowledgments}

\bibliography{IEEEabrv,bib.bib}

\end{document}